\documentclass[aip,graphicx]{revtex4-1}
\draft % marks overfull lines with a black rule on the right
\usepackage{lineno,hyperref}
\usepackage{subfigure}
\usepackage{graphicx}
\usepackage{epstopdf}
\usepackage{mathrsfs}
\begin{document}

%\preprint{}

\title{Spectral Camera based on true thermal light Ghost Imaging via Sparsity Constraints} %Title of paper



\author{Zhentao Liu}
\author{Shiyu Tan}
\author{Jianrong Wu}
\author{Enrong Li}
\author{Shensheng Han}
\email[]{sshan@mail.shcnc.ac.cn}
%\thanks{}
%\author{Li Guo}
%\author{Shensheng Han}
\affiliation{Key Laboratory for Quantum Optics and Center for Cold Atom Physics, Shanghai Institute of Optics and Fine Mechanics, Chinese Academy of Sciences, Shanghai 201800, China}


\date{\today}

\begin{abstract}
The image information acquisition ability of a conventional camera is usually much lower than the Shannon Limit since it does not make use of the correlation between pixels of image data. Applying a random phase modulator to code the spectral images and combining with compressive sensing (CS) theory, a spectral camera based on true thermal light ghost imaging via sparsity constraints (GISC spectral camera) is proposed and demonstrated experimentally. GISC spectral camera can acquire the information at a rate significantly below the Nyquist rate, and the resolution of the cells in the three-dimensional (3D) spectral images data-cube can be achieved with a two-dimensional (2D) detector in a single exposure. For the first time, GISC spectral camera opens the way of approaching the Shannon Limit determined by Information Theory in optical imaging instruments.
\end{abstract}


\maketitle %\maketitle must follow title, authors, abstract and \pacs

\section{Introduction}
Conventional Camera, as one of the most important appliances to get image information, records the image of an object based on the point-to-point correspondence between the object-space and the image-space. Because the correlation between pixels of image \cite{jacobs1992image}can¡¯t be applied, the image information acquisition efficiency of such conventional point-to-point imaging mode is much lower than the Shannon Limit\cite{SHANNON1948Communication, cover2012elements} determined by Information Theory in optical imaging instruments\cite{elias1953optics, francia1955resolving, Tan1964Optical, di1969degrees, Tan1982Optical, huck1999information, strange2005information}. Unlike the conventional direct point-to-point imaging mode, the resolution of the pixels of ghost imaging is determined by the correlation of light field fluctuations corresponding to the two pixels respectively, which can be measured on-line or pre-determined.\cite{kolobov2007quantum, shapiro2012physics} Combining with compressive sensing (CS) theory\cite{jacobs1992image, gonzalez2004digital, donoho2006compressed, candes2006robust, eldar2012compressed, wu2014snapshot}, ghost imaging via sparsity constraints (GISC) has many potential applications including super-resolution imaging\cite{gong2009super, gong2012experimental, wang2012quantum, gong2015high}, three-dimensional (3D) computational imaging with single-pixel detectors\cite{sun20133d}, 3D remote sensing\cite{zhao2012ghost, gong2013three}, imaging through scattering media\cite{gong2011correlated, bina2013backscattering}, object tracking\cite{magana2013compressive}, object authentication\cite{chen2013object,xu2014morphology} and X-ray Fourier transform diffraction imaging\cite{cheng2004incoherent, zhang2007lensless, wang2012coherent}.
For thermal light ghost imaging, according to the illumination source, it can be classified to two categories: ghost imaging with pseudo-thermal light and true thermal light. Ghost imaging with true thermal light and sunlight have been respectively demonstrated by detecting the temporal fluctuation of thermal light and applying the intensity correlation between the intensity distributions at the reference arm and the test arm.\cite{zhang2005correlated, d2005quantum, liu2014lensless} Comparing with ghost imaging with pseudo-thermal light, this scheme of ghost imaging with true thermal light has to face the difficulty of detecting the temporal fluctuation of true thermal light which requires the response time of detector less than the coherence time of true thermal light $\tau  = \frac{{{{{\lambda ^2}} \mathord{\left/
 {\vphantom {{{\lambda ^2}} {\Delta \lambda }}} \right.
 \kern-\nulldelimiterspace} {\Delta \lambda }}}}{c} \propto \frac{1}{{\Delta \lambda }}$  ( $\lambda $ is the wavelength,  ${\Delta \lambda }$ is the linewidth,  $c$ is the speed of light) which can be as short as femtosecond. In order to increase the coherence time of the illumining true thermal light, monochrome imaging is required which results in the vast majority of radiation energy from the target scene being filtered out, making the energy efficiency of ghost imaging applying the temporal fluctuation of true thermal light very low. Moreover, the fluctuating true thermal field needs to be split before the light field illuminating the object in the system and recorded in the reference path£¬which makes the scheme even more difficult to be applied in remote sensing.
 In this paper, for the first time, we propose a spectral camera based on true thermal light ghost imaging via sparsity constraints (GISC spectral camera) without a splitter. GISC spectral camera modulates the true thermal light into a spatially fluctuating pseudo-thermal light using a spatial random phase modulator\cite{giglio2000space, cerbino2008x} which, at the same time, also acts as a random grating generating the uncorrelated speckles for different wavelengths, the 3D spectral images data-cube is then modulated into a two-dimensional (2D) data plane and GISC spectral camera can achieve the whole wavelength image in a single exposure, leading to a more convenient detection process and higher energy efficiency compared to ghost imaging applying the temporal fluctuation of true thermal light. Combining with CS, GISC spectral camera can acquire the information at a rate significantly below the Nyquist rate which opens the way of approaching the Shannon Limit determined by Information Theory in optical imaging instruments\cite{cover2012elements, elias1953optics, francia1955resolving, Tan1964Optical, Tan1982Optical}.

\section{Schematic \& Resolution}
\begin{figure}[ht]
\centering
\includegraphics[width=10.0cm]{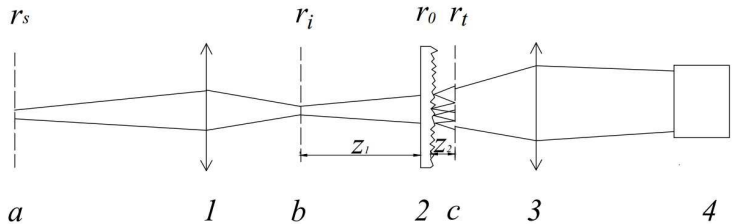}%
\caption{Schematic of GISC spectral camera. ($a$) The object plane; ($b$) the first image plane; ($c$) the speckles plane; (1) an imaging system; (2) a spatial random phase modulator; (3) a microscope objective; (4) CCD detector.}%
\label{fig1}
\end{figure}

The schematic of GISC spectral camera is shown in Fig.\ref{fig1}. The system consists of (1) an imaging system, which projects the object image in the object plane `$a$' onto the first image plane `$b$', (2) a spatial random phase modulator, which disperses the image with different wavelengths as a random grating and modulates the image to generate the speckles in plane `$c$'\cite{giglio2000space, cerbino2008x}, (3) a microscope objective, which magnifies the speckles in plane `$c$', and (4) a charge-coupled device (CCD) detector recording the magnified speckles.

\begin{figure}[ht]
  \centering
  \subfigure[]{
    \label{subfig2a} %% label for first subfigure
    \includegraphics[width=5cm]{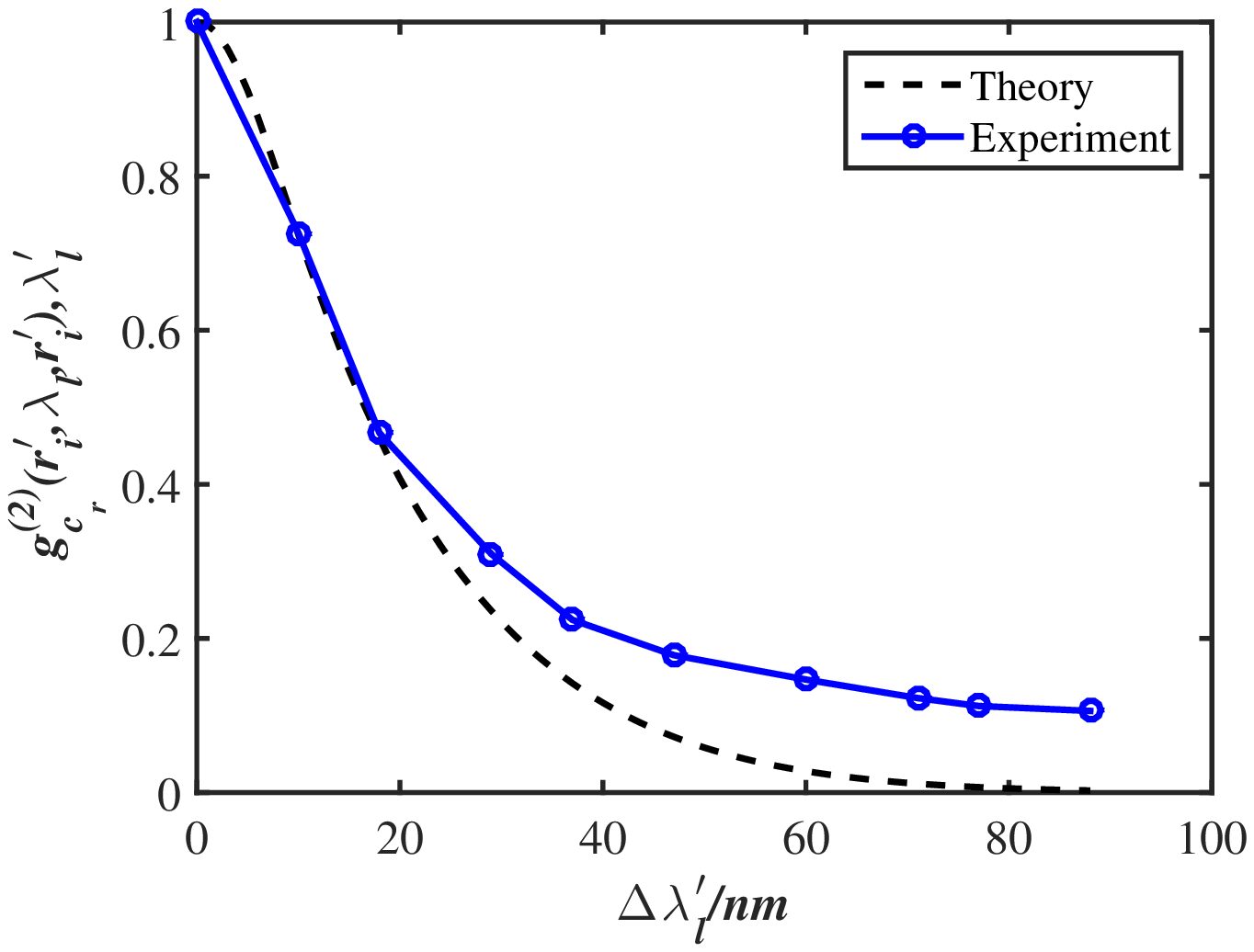}}
  \hspace{1in}
  \subfigure[]{
    \label{subfig2b} %% label for second subfigure
    \includegraphics[width=5cm]{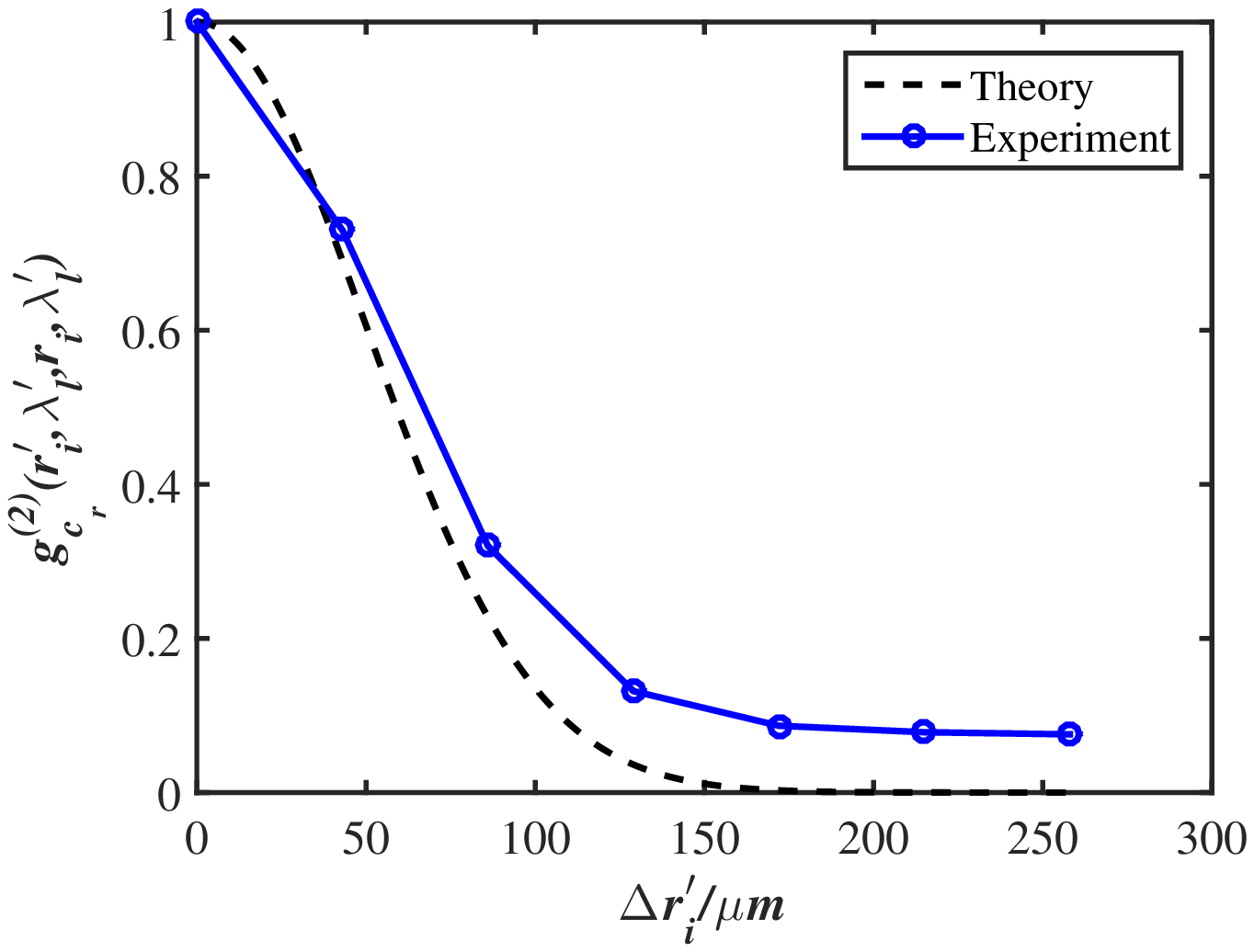}}
  \caption{(a) The normalized second-order correlation function of the light fields $\emph{g}_{c_r}^{(2)}(r_i',\lambda_l,r_i',\lambda_l')$ at pixel   in the first image plane `$b$' with two different wavelengths; (b) The normalized second-order correlation function of the light fields $\emph{g}_{c_r}^{(2)}(r_i,\lambda_l',r_i',\lambda_l')$ at two different pixels in the first image plane `$b$' with wavelength $\lambda_l'$.}
  \label{fig2} %% label for entire figure
\end{figure}

Denoting the spectral light intensity distribution in the first image plane `$b$' by ${I_b}\left( {{r_i},{\lambda _l}} \right)$ and the intensity distribution in plane `$c$' by ${I_c}\left( {{r_t}} \right)$ respectively, we have\cite{goodman2005introduction}
\begin{equation}\label{eq:1}
  I_c(r_t) = \int\int I_b(r_i,\lambda_l)h_I(r_t;r_i,\lambda_l)dr_id\lambda_l,
\end{equation}
where $h_I(r_t;r_i,\lambda_l)$ is the incoherent intensity impulse response function, $r_t$ is the coordinate in plane `$c$', $r_i$ and $\lambda_l$ are respectively the coordinate and wavelength of the light intensity distribution in the first image plane `$b$'.
To record the pre-determined reference spatial intensity fluctuation of the pseudo-thermal light without objects, a coherent monochrome point source at pixel $r_i'$ with wavelength $\lambda_l'$ in the first image plane `$b$', denoted as $I_{b_r}(r_i,\lambda_l;r_i',\lambda_l')=\delta(r_i-r_i',\lambda_l-\lambda_l')$, is used to illuminate the spatial random phase modulator, and the recorded light intensity $I_{c_r}(r_t;r_i',\lambda_l')$ in the plane `$c$' is given by
\begin{eqnarray}\label{eq:2}
   \nonumber I_{c_r}(r_t;r_i',\lambda_l') &=& \int\int
  I_{b_r}(r_i,\lambda_l)h_I(r_t;r_i,\lambda_l)dr_id\lambda_l \\
   &=& \int\int
  \delta(r_i-r_i',\lambda_l-\lambda_l')h_I(r_t;r_i,\lambda_l)dr_id\lambda_l \\
   \nonumber &=& h_I(r_t;r_i',\lambda_l').
\end{eqnarray}
During the imaging process, the intensity distribution in the first image plane `$b$' $I_{b_t}(r_i,\lambda_l)$ is simply the image, denoted as $T_i(r_i,\lambda_l)$, of the object $T_s(r_s,\lambda_l)$ in the object plane `$a$',
\begin{equation}\label{eq:3}
  I_{b_t}(r_i,\lambda_l)=T_i(r_i,\lambda_l).
\end{equation}
Combining Eqs. (\ref{eq:1})(\ref{eq:2}) with (\ref{eq:3}), the intensity distribution $I_{c_t}(r_t)$ in the speckle plane `$c$' is
\begin{equation}\label{eq:4}
  I_{c_t}(r_t)=\int\int
  T_i(r_i,\lambda_l)I_{c_r}(r_t;r_i,\lambda_l)dr_id\lambda_l.
\end{equation}
Eq. (\ref{eq:4}) shows that $I_{c_t}(r_t)$ is the $T_i(r_i,\lambda_l)$ weighted integration of the pre-determined reference spatial intensity fluctuation of pseudo-thermal light $I_{c_r}(r_t;r_i,\lambda_l)$. Therefore, each pixel $r_t$ of CCD detector is equivalent to a measurement of the bucket detector in the test arm of ghost imaging scheme.
The second-order correlation function between the spatial intensity fluctuation in the pre-determined reference arm and test arm is defined as
\begin{equation}\label{eq:5}
  G^{(2)}(r_i',\lambda_l')=\langle{E_{c_r}^*(r_t;r_i',\lambda_l')E_{c_t}^*(r_t)E_{c_t}(r_t)E_{c_r}(r_t;r_i',\lambda_l')}\rangle_{r_t},
\end{equation}
where $\langle{...}\rangle_{r_t}$ is the ensemble average about the coordinate of the light intensity distribution $r_t$. Combining Eqs. (\ref{eq:2})(\ref{eq:4}) with (\ref{eq:5}), the second-order correlation function $G^{(2)}(r_i',\lambda_l')$ is given by
\begin{equation}\label{eq:6}
  G^{(2)}(r_i',\lambda_l')=\int\int
  T_i(r_i,\lambda_l)G_{c_r}^{(2)}(r_i,\lambda_l,r_i',\lambda_l')dr_id\lambda_l,
\end{equation}
where $G_{c_r}^{(2)}(r_i,\lambda_l,r_i',\lambda_l')=\langle{E_{c_r}^*(r_t;r_i',\lambda_l')E_{c_r}^*(r_t;r_i,\lambda_l)E_{c_r}(r_t;r_i,\lambda_l)E_{c_r}(r_t;r_i',\lambda_l')}\rangle_{r_t}$ is the second-order correlation function of the light fields at different pixels and wavelengths in the first image plane `$b$'.
In order to calculate $G_{c_r}^{(2)}(r_i,\lambda_l,r_i',\lambda_l')$, the height autocorrelation function of the spatial random phase modulator is assumed as\cite{cheng1999computational}
\begin{equation}\label{eq:7}
  R_\eta(r_0,r_0')=\langle{\eta(r_0)\eta(r_0')}\rangle=\omega^2exp{-{\frac{r_0-r_o'}{\zeta}}^2}=R_\eta(\Delta{r_0}),\Delta(r_0)=r_0-r_0',
\end{equation}
where $\eta(r_0)$ and $\eta(r_0')$ are respectively the height of the spatial random phase modulator at $r_0$ and $r_0'$, $\omega$ and $\zeta$ are respectively the height standard deviation and transverse correlation length of the spatial random phase modulator. Assuming that the light field fluctuations in the speckles plane `$c$' corresponding to pixel $r_i'$ in the first image plane `$b$' with wavelength $\lambda_l'$ obeys the complex circular Gaussian distribution, $G_{c_r}^{(2)}(r_i,\lambda_l,r_i',\lambda_l')$ can be written as
\begin{eqnarray}\label{eq:8}
  G_{c_r}^{(2)}(r_i,\lambda_l,r_i',\lambda_l') \nonumber &=& \langle{E_{c_r}^*(r_t;r_i',\lambda_l')E_{c_r}^*(r_t;r_i,\lambda_l)E_{c_r}(r_t;r_i,\lambda_l)E_{c_r}(r_t;r_i',\lambda_l')}\rangle_{r_t} \\
   \nonumber &=& \langle{E_{c_r}^*(r_t;r_i',\lambda_l')E_{c_r}(r_t;r_i',\lambda_l')}\rangle_{r_t} \langle{E_{c_r}^*(r_t;r_i,\lambda_l)E_{c_r}(r_t;r_i,\lambda_l)}\rangle_{r_t}
  \\  &+&\langle{E_{c_r}^*(r_t;r_i',\lambda_l')E_{c_r}^*(r_t;r_i,\lambda_l)}\rangle_{r_t} \langle{E_{c_r}(r_t;r_i,\lambda_l)E_{c_r}(r_t;r_i',\lambda_l')}\rangle_{r_t}  \\
   \nonumber &=& \langle{I_{c_r}(r_t;r_i',\lambda_l')}\rangle_{r_t} \langle{I_{c_r}(r_t;r_i,\lambda_l)}\rangle_{r_t}+{\mid{J_{c_r}(r_i,\lambda_l,r_i',\lambda_l')}\mid}^2 \\
   \nonumber &=& \langle{I_{c_r}(r_t;r_i',\lambda_l')}\rangle_{r_t} \langle{I_{c_r}(r_t;r_i,\lambda_l)}\rangle_{r_t}[1+\emph{g}_{c_r}^{(2)}((r_i,\lambda_l,r_i',\lambda_l')],
\end{eqnarray}
where
\begin{equation}\label{eq:9}
  J_{c_r}(r_i,\lambda_l,r_i',\lambda_l')=\langle{E_{c_r}^*(r_t;r_i',\lambda_l')E_{c_r}^*(r_t;r_i,\lambda_l)}\rangle_{r_t},
\end{equation}
\begin{equation}\label{eq:10}
  \emph{g}_{c_r}^{(2)}((r_i,\lambda_l,r_i',\lambda_l')=\frac{{\mid{J_{c_r}(r_i,\lambda_l,r_i',\lambda_l')}\mid}^2 }{\langle{I_{c_r}(r_t;r_i',\lambda_l')}\rangle_{r_t} \langle{I_{c_r}(r_t;r_i,\lambda_l)}\rangle_{r_t}}.
\end{equation}
$\emph{g}_{c_r}^{(2)}((r_i,\lambda_l,r_i',\lambda_l')$ is defined as the normalized second-order correlation function of the light fields at different pixels and wavelengths in the first image plane `$b$'.
According to the Fresnel diffraction theorem, the light field in the speckles plane `$c$' propagated from pixel $r_i'$ in the first image plane `$b$' with wavelength $\lambda_l'$ is
\begin{eqnarray}\label{eq:11}
  E_{c_r}(r_t;r_i',\lambda_l') \nonumber &=& \frac{exp\{{j\pi[2(z_1+z_2)+(r_t-r_i')^2/(z_1+z2)]/{\lambda_l'}}\}}{j\lambda_l'z_1z_2} \\
   &\times& \int
   t(r_0,\lambda_l')exp[j\frac{\pi(z_1+z_2)}{\lambda_l'z_1z_2}(r_0-\frac{z_1r_t+z_2r_i'}{z_1+z_2})^2]dr_0,
\end{eqnarray}
where $t(r_0,\lambda_l')=exp[j2\pi(n-1){\eta(r_0)}/{\lambda_l'}]$ is the transmission function of the spatial random phase modulator. $\langle{I_{c_r}(r_t;r_i',\lambda_l')}\rangle$ and $\langle{I_{c_r}(r_t;r_i,\lambda_l)}\rangle$ are respectively given by
\begin{equation}\label{eq:12}
  \langle{I_{c_r}(r_t;r_i',\lambda_l')}\rangle_{r_t}=\langle{E_{c_r}^*(r_t;r_i',\lambda_l')E_{c_r}(r_t;r_i',\lambda_l')}\rangle_{r_t} =\frac{1}{z_1z_2\lambda_l'(z_1+z_2)},
\end{equation}
\begin{equation}\label{eq:13}
  \langle{I_{c_r}(r_t;r_i,\lambda_l)}\rangle_{r_t}=\langle{E_{c_r}^*(r_t;r_i,\lambda_l)E_{c_r}(r_t;r_i,\lambda_l)}\rangle_{r_t} =\frac{1}{z_1z_2\lambda_l(z_1+z_2)}.
\end{equation}
Substituting Eqs. (\ref{eq:7})(\ref{eq:9})(\ref{eq:11}) into (\ref{eq:10}) yields
\begin{eqnarray}\label{eq:14}
  \emph{g}_{c_r}^{(2)}((r_i,\lambda_l,r_i',\lambda_l') &=& \frac{(z_1+z_2)^2exp\{-[2\pi\omega(n-1)]^2(\frac{1}{\lambda_l^2}+\frac{1}{\lambda_l'^2})\}}{z_1^2z_2^2\lambda_l\lambda_l'} \\
   \nonumber &\times& |\int\int exp\{[2\pi(n-1)]^2\frac{R_\eta(\mu,\nu)}{\lambda_l\lambda_l'}\}exp\{j\frac{\pi(z_1+z_2)}{z_1z_2}(\frac{\mu^2}{\lambda_l}-\frac{\nu^2}{\lambda_l'})\}d\mu d\nu|^2,
\end{eqnarray}
where
\begin{equation}\label{eq:15}
  \mu=r_0-\frac{z_1r_t+z_2r_i}{z_1+z_2},
\end{equation}
\begin{equation}\label{eq:16}
  \nu=r_0'-\frac{z_1r_t+z_2r_i'}{z_1+z_2},
\end{equation}
\begin{equation}\label{eq:17}
  \Delta\lambda_l'=\lambda_l'-\lambda_l,
\end{equation}
\begin{equation}\label{eq:18}
  R_\eta(\mu,\nu)=\omega^2exp\{-[(\mu-\nu)-\frac{z_2(r_i-r_i')}{z_1+z_2}/\zeta]^2\}=\omega^2exp\{-[(\mu-\nu)-\frac{z_2\Delta r_i'}{z_1+z_2}/\zeta]^2\},
\end{equation}
\begin{equation}\label{eq:19}
  \Delta r_i'=r_i-r_i'.
\end{equation}
Assuming $\frac{|\Delta\lambda_l'|}{\lambda_l}\ll1$, we have
\begin{eqnarray}\label{eq:20}
  \nonumber &g&_{c_r}^{(2)}((r_i,\lambda_l,r_i',\lambda_l')\\
   &\approx& \frac{(z_1+z_2)^2exp\{-[2\pi\omega(n-1)]^2(\frac{1}{\lambda_l^2}+\frac{1}{\lambda_l'^2})\}}{z_1^2z_2^2\lambda_l\lambda_l'} \\
   \nonumber &\times& |\int\int exp\{[2\pi(n-1)]^2\frac{R_\eta(\mu,\nu)}{\lambda_l\lambda_l'}\}exp\{j\frac{\pi(z_1+z_2)}{z_1z_2\lambda_l}(\mu^2-\nu^2+\frac{\Delta\lambda_l'}{\lambda_l}\nu^2)\}d\mu d\nu|^2.
\end{eqnarray}
Assuming $\frac{z_1}{z_2}\gg1$, and the diameter $\sigma$ of the illuminated region in the spatial random phase modulator by each cells of 3D data-cube in calibration satisfies ${\pi\sigma^2}/{\lambda_lz_2}<1$, $\emph{g}_{c_r}^{(2)}((r_i,\lambda_l,r_i',\lambda_l')$ is given by
\begin{equation}\label{eq:21}
  \emph{g}_{c_r}^{(2)}((r_i,\lambda_l,r_i',\lambda_l')\approx exp\{-[2\pi\omega(n-1)]^2(\frac{1}{\lambda_l}-\frac{1}{\lambda_l'})^2\}exp\{\frac{2[2\pi\omega(n-1)]^2}{\lambda_l^2}\{exp\{-[\frac{z_2\Delta r_i'}{(z_1+z_2)\zeta}]^2\}-1\}\}.
\end{equation}
Taking Eqs. (\ref{eq:8})(\ref{eq:12})(\ref{eq:13}) and (\ref{eq:21}) into Eq. (\ref{eq:6}), we get the correlation function of intensity fluctuations\cite{gatti2004ghost}
\begin{eqnarray}\label{eq:22}
  \nonumber &\Delta& G^{(2)}(r_i',\lambda_l') \\
  \nonumber &=& G^{(2)}(r_i',\lambda_l')-\langle{I_{c_r}(r_t;r_i,\lambda_l)}\rangle_{r_t}\langle{I_{c_t}(r_t)}\rangle_{r_t}  \\
  \nonumber &\approx& -\frac{1}{z_1^2z_2^2(z_1+z_2)^2}\{\{T_i(r_i',k_l')\}_{r_i'}\otimes\{exp\{2[2\pi\omega(n-1)k_l']^2\{exp\{-[\frac{z_2r_i'}{(z_1+z_2)\zeta}]^2\}-1\}\}\}_{r_i'}\}_{k_l'} \\
  &\otimes& exp\{-[2\pi\omega(n-1)]^2k_l'^2\},
\end{eqnarray}
where $k_l=\frac{1}{\lambda_l}$, $k_l'=\frac{1}{\lambda_l'}$, $\otimes$ denotes the operation of convolution. Eq. (\ref{eq:22}) specifies that $T_i(r_i',\lambda_l')$ can be separated from the correlation function of intensity fluctuations $\Delta G^{(2)}(r_i',\lambda_l')$, and the resolution is determined by the normalized second-order correlation $g_{c_r}^{(2)}((r_i,\lambda_l,r_i',\lambda_l')$ at different pixels and wavelengths in the first image plane `$b$'.
When $r_i=r_i'$, according to Eq. (\ref{eq:14}), the normalized second-order correlation function of the light fields at pixel $r_i'$ in the first image plane `$b$' with two different wavelengths is given by
\begin{eqnarray}\label{eq:23}
  \nonumber &\emph{g}&_{c_r}^{(2)}(r_i',\lambda_l,r_i',\lambda_l')  \\
  &=&\frac{(z_1+z_2)^2exp\{-[2\pi\omega(n-1)]^2(\frac{1}{\lambda_l^2}+\frac{1}{\lambda_l'^2})\}}{z_1^2z_2^2\lambda_l\lambda_l'}\times|\int\int exp\{[2\pi(n-1)]^2\frac{R_\eta(r_0,r_0')}{\lambda_l\lambda_l'}\}\\
  \nonumber &exp&\{j\frac{\pi(z_1+z_2)}{z_1z_2}[(r_0-\frac{z_1r_t+z_2r_i}{z_1+z_2})^2/\lambda_l-(r_0'-\frac{z_1r_t+z_2r_i'}{z_1+z_2})^2/\lambda_l']\}dr_0 dr_0'|^2.
\end{eqnarray}
Similarly, when $\lambda_l=\lambda_l'$, according to Eq. (\ref{eq:14}), the normalized second-order correlation function of the light fields at two different pixels in the first image plane `$b$' with wavelength $\lambda_l'$ is given by
\begin{equation}\label{eq:24}
 \emph{g}_{c_r}^{(2)}(r_i,\lambda_l',r_i',\lambda_l')=exp\{-2[2\pi\omega(n-1)/\lambda_l']^2\{1-exp[-(\frac{z_2\Delta r_i'}{(z_1+z_2)\zeta})^2]\}\}.
\end{equation}
Fig.\ref{subfig2a} and Fig.\ref{subfig2b} respectively show the comparison of $\emph{g}_{c_r}^{(2)}(r_i',\lambda_l,r_i',\lambda_l')$ and $\emph{g}_{c_r}^{(2)}(r_i,\lambda_l',r_i',\lambda_l')$ between experiment and theory, and the experiment diagram is given in Fig.\ref{fig1} with $z_1=20mm$, $z_2=0.3mm$, $\omega=2.1\mu m$, $\zeta=16.75\mu m$, $n=1.516$ and the central wavelength $\lambda_l'=600nm$.

\section{The Measurement Matrix \& Reconstruction Algorithm.}
There are many methods to improve the imaging quality of ghost imaging.\cite{chan2009high, gong2010method, ferri2010differential} However, ghost imaging reconstructions based on the ensemble statistics cannot provide the criterion of the necessary number of sampling for a perfect imaging, which makes it impossible to optimize the design of ghost imaging system. Combining with CS which provides the recovery condition of perfect reconstruction, the quantitative analysis for the necessary measurements data can be made. Under the framework of CS theory, the measurement matrix of GISC spectral camera is obtained as follows: each of the speckle intensity distributions generated by a point light source at pixel $r_i'$ in the $\lambda\pm\Delta\lambda_l'$ spectrum band in the first image plane `$b$' is recorded by the randomly selected $M_{r_t}$ pixels of CCD detector and reshaped as a column vector of length $M$ of the measurement matrix. Repeating the process for all the $N$ image pixels in the first image plane `$b$' and all the $L$ spectral bands, one may have the pre-determined random measurement matrix $A_{M\times K}$, where $K=L\times N$. If we denote the unknown spectral object image as a $K$-dimensional column vector $X_{K\times 1}$, and reshape the modulated object intensity distribution recorded by the same $M$ pixels of CCD detector in a similar way as a column vector $Y_{M\times 1}$, then we may have the discrete from Eq. (\ref{eq:4}),
\begin{equation}\label{eq:25}
  Y=AX.
\end{equation}
Spectral object image is usually both spatially and spectrally correlated, which has already been utilized in spectral image reconstructions.\cite{oymak2012simultaneously, golbabaee2012compressed, golbabaee2012joint} The reconstruction of the spectral object image can generally be regarded as solving a minimization problem which penalizes both the $l_1$ norm and the nuclear norm of the data matrix:
\begin{equation}\label{eq:26}
 \mathop{\min}\limits_{X}\mu_1\|\psi X\|_{l_1}+\mu_2\|{\tilde{X}}\|_*,s.t.,Y=AX,
\end{equation}
where $\tilde{X}_{L\times N}$ a matrix representation of the spectral object image whose columns represent different bands of the spectral object image, $\psi$ the sparsifying transform, $\mu_1$ and $\mu_2$ the weight coefficients and $\mu_1,\mu_2>0$. In this work, we use a modified approach based on the method described by Eq. (\ref{eq:26})\cite{zhang2014hyperspectral}:
\begin{equation}\label{eq:27}
 \mathop{\min}\limits_{X}\mu_1\|\psi X\|_{l_1}+\mu_2\|\Delta s\|_{l_1},s.t.,Y=AX,X\geq0,
\end{equation}
where $\Delta s_i=s_{i+1}-s_1$, is the subtraction of the largest singular value $s_1$ and the other $s_i$. The solution of Eq. (\ref{eq:27}) tends to have a simultaneous low-effective-rank and sparse structure, which much improves the reconstruction quality with low sampling rate.
\section{Experimental Results..}
\begin{figure}[ht]
\centering
\includegraphics[width=10.0cm]{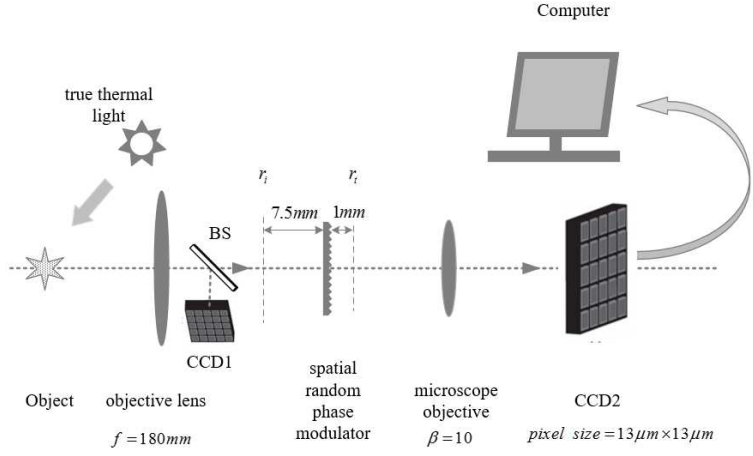}%
\caption{Experimental setup of GISC spectral camera.}%
\label{fig3}
\end{figure}

\begin{figure}[ht]
\centering
\includegraphics[width=15.0cm]{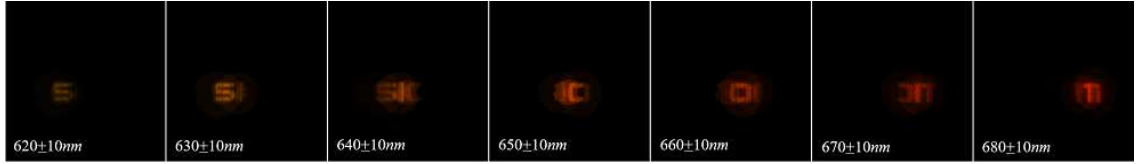}%
\caption{The original spectral images of `SIOM' obtained by CCD1 detector placed on the first image plane `$b$' with corresponding narrowband filter in front of it, showing all the channels from $620\sim680 nm$.}%
\label{fig4}
\end{figure}

\begin{figure}[ht]
\centering
\includegraphics[width=15.0cm]{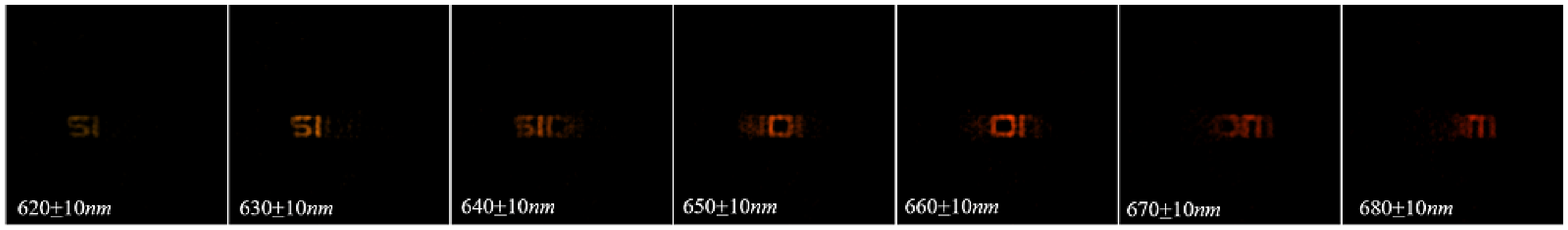}%
\caption{The reconstructed spectral images of `SIOM' with $30\%$ sampling rate of 3D date-cube, showing all the channels from $620\sim680nm$.}%
\label{fig5}
\end{figure}

In the experimental setup of GISC spectral camera shown in Fig.\ref{fig3}, the imaging system (Tamron AF70-300$mm$ f/4-5.6) with focal length of $f=180mm$ projects the object image onto the first image plane, a beam splitter (BS) with split ratio 50:50 splits the light field into two paths, CCD1 detector (AVT Sting F-504C with pixel size of $3.45\mu m\times 3.45\mu m$) is placed in one of the two paths at the position of the first image plane of the system to obtain the conventional image of the object for comparison, a spatial random phase modulator (SIGMA KOKI CO., LTD. DFSQ1-30C02-1000) disperses the images with different wavelengths acting as a random grating and modulates the image to generate the speckles, a microscope objective with magnification $\beta=10$ and the numerical aperture $N.A.=0.25$ magnifies the speckles which are then recorded by CCD2 detector (Andor iKon-M) with the pixel size $13\mu m\times13\mu m$. The first image plane is divided into $N_x\times N_y=140 \times140$ pixels with the square of each pixel approximately equal to $\Delta r_s$ determined by the Eq. (\ref{eq:24}). The number of spectrum bands for single exposure is 7, and the images in two wavelength ranges of $520\sim580 nm$ and $620\sim680 nm$ are respectively obtained in two exposures, while the theoretical spectral resolution is $20 nm$ in the experimental setup according to Eq. (\ref{eq:23}).
In order to compare the spectral \& spatial resolution of GISC spectral camera with the theoretical resolution, as shown in Fig.\ref{fig4}, the spectral object `SIOM' with different parts passing through different wavelengths has been selected, and the illuminating source is a xenon lamp. The original spectral images of `SIOM' obtained by CCD1 detector placed in the first image plane `$b$' with corresponding narrowband filter in front of it are shown in Fig.\ref{fig4} (pixel size is equal to the theoretical resolution of reconstructed images by GISC spectral camera for comparing them). The corresponding modulated object intensity distribution $Y$ is achieved by CCD2 detector of GISC spectral camera and the reconstructed spectral images of `SIOM' with $30\%$ sampling rate of 3D date-cube are shown in Fig.\ref{fig5}. The comparison between the original and reconstructed spectral images shows that the resolution of GISC spectral camera is in accordance with the theoretical calculation.

\begin{figure}[ht]
 \centering
 \subfigure[]{
  \label{subfig6a} %% label for first subfigure
  \includegraphics[width=2.5cm]{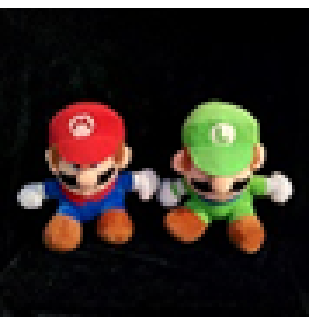}}
  \subfigure[]{
  \label{subfig6b} %% label for first subfigure
  \includegraphics[width=2.5cm]{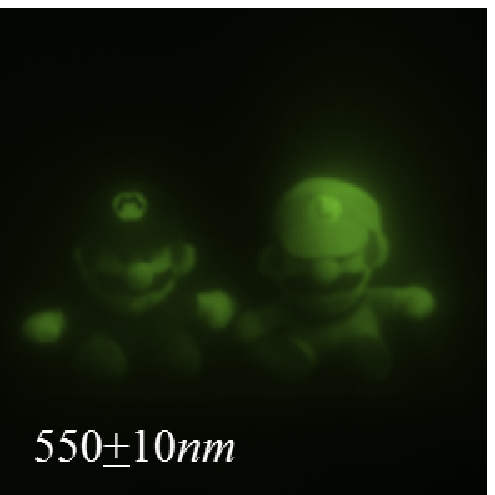}}
  \subfigure[]{
    \label{subfig6c} %% label for first subfigure
    \includegraphics[width=2.5cm]{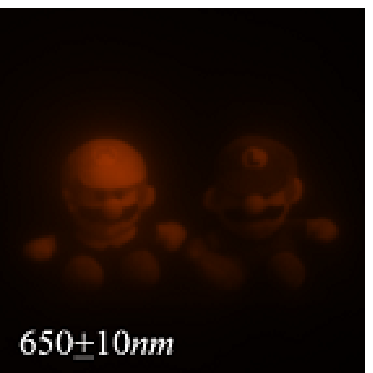}}
  \caption{taken by (a) a camera; (b) CCD1 detector passing through narrowband filters of $550\pm10nm$; (c) CCD1 detector passing through narrowband filters of $650\pm10nm$.}
  \label{fig6} %% label for entire figure
\end{figure}

\begin{figure}[ht]
\centering
\includegraphics[width=15.0cm]{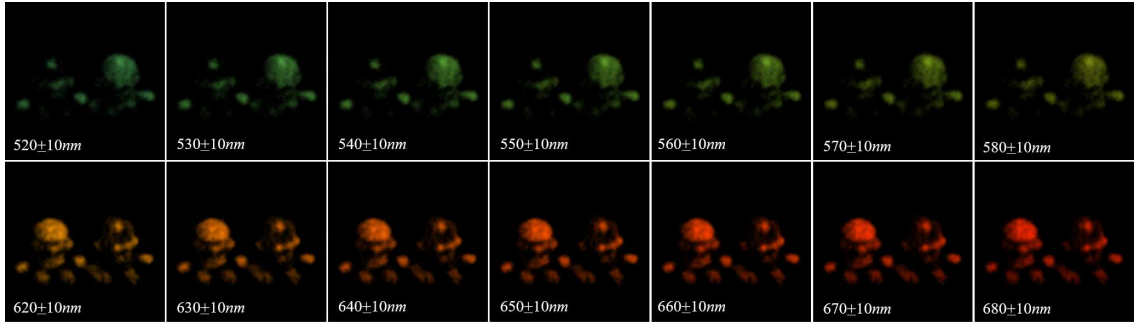}%
\caption{The reconstructed spectral images of Mario \& Luigi with $30\%$ sampling rate of 3D date-cube, showing all the channels from $520\sim580 nm$ and $620\sim680 nm$.}%
\label{fig7}
\end{figure}

The images of the outdoor scene consisting of Mario \& Luigi with sunlight illumination are shown in Fig.\ref{fig6}. Fig.\ref{subfig6a} is obtained by a camera, while Fig.\ref{subfig6b} and Fig.\ref{subfig6c} respectively show the pictures taken by CCD1 detector with narrowband filters of $550\pm 10nm$ and $650\pm10 nm$ in front of it (pixel size is equal to the theoretical resolution of reconstructed images by GISC spectral camera for the sake of comparison). The reconstructed spectral images of Mario \& Luigi with $30\%$ sampling rate of 3D date-cube are shown in Fig.\ref{fig7}. The experimental results show that the spectral imaging ability of GISC spectral camera for complex scenes is also pretty good.

\section{Discussion \& Conclusion.}
Based on Information Theory, the transmitted information of an imaging system can be described by the entropy\cite{SHANNON1948Communication, cover2012elements}
\begin{equation}\label{eq:28}
  H(X)=-\mathop{\sum}\limits_{i}p(x_i)\log p(x_i),
\end{equation}
where $p(x_i)$ is the probability of $x_i$ occurrence. For the conventional direct point-of-object-space to point-of-image-space imaging mode, the conditional entropy $H(X|Y)=0$, and thus the channel capacity of the conventional monochrome camera is
\begin{equation}\label{eq:29}
  C=\mathop{\max}\limits_{p(x_i)}I(X;Y)=\mathop{\max}\limits_{p(x_i)}[H(X)-H(X|Y)]=\mathop{\max}\limits_{p(x_i)}H(X)=H_C(X),
\end{equation}
where $I(X;Y)$ is the mutual information, $H_C(X)= \mathop{\max}\limits_{p(x_i)}H(X)$ the maximum information entropy of source $X$ for conventional imaging instrument, which is the Shannon Limit of the imaging system. According to the principle of maximum entropy\cite{jacobs1992image}, the information content of an image is maximized when $p(x_i)$ is Gaussian distribution with average power constraints, which doesn¡¯t contain any useful information. Therefore, the entropy of the image with structured information $H(X)$ has
\begin{equation}\label{eq:30}
  H(X)<H_C(X)=C.
\end{equation}
Eq. (\ref{eq:30}) shows that the image information acquisition efficiency of such conventional point-to-point imaging mode is lower than the Shannon Limit determined by Information Theory in optical imaging instruments.
The channel capacity of an imaging system based on Information Theory for conventional optical imaging instruments is\cite{elias1953optics, francia1955resolving, Tan1964Optical, di1969degrees, Tan1982Optical}
\begin{equation}\label{eq:31}
  C=N_{DOF}\log_2(1+m),
\end{equation}
where $m$ is signal to noise ratio (SNR), $N_{DOF}$ is degrees of freedom and has
\begin{equation}\label{eq:32}
  N_{DOF}=N_t\cdot N_s\cdot N_c\cdot N_\phi,
\end{equation}
where $N_t,N_s,N_c$ and $N_\phi$ are respectively time, spatial, color and polarization degrees of freedom.
Spatial degrees of freedom $N_s$ has\cite{di1969degrees}
\begin{equation}\label{eq:33}
  N_s=SW=(N_x\frac{0.61\lambda}{\alpha_x})(N_y\frac{0.61\lambda}{\alpha_y})\frac{1}{\lambda/(2\alpha_x)}frac{1}{\lambda/(2\alpha_y)}=1.22^2N_xN_y,
\end{equation}
where $S$ is the image area, $W$ is the space bandwidth, $\alpha_x$, $\alpha_y$ and $N_x$, $N_y$ are respectively the image-space aperture angle and the resolved pixel number in the image-space of coordinate $x$ and $y$. The color degrees of freedom $N_c$ depend on the number of spectral channels, while polarization degrees of freedom $N_\phi$ is determined by the independent polarization state.
According to Eqs. (\ref{eq:28})(\ref{eq:31}) and (\ref{eq:33}), the channel capacity of the conventional camera in our experiment (where $N_{t_1}=1, N_{c_1}=1, N_{\phi_1}=1, N_{x_1}\times N_{y_1}=140\times 140, m_1=255$) is $C_1=N_{DOF_1}\log_2(1+m_1)\approx 2.33\times10^5$, and the corresponding transmitted information of Fig.\ref{subfig6b} is $H_1(X)\approx1.07\times10^5<C_1$. In order to transmit the $520\sim580nm$ wavelength ranges data, the required channel capacity of the conventional camera (where $N_{t_2}=1, N_{c_2}=7, N_{\phi_2}=1, N_{x_2}\times N_{y_2}=140\times 140, m_2=255$) is $C\approx1.63\times10^6$, while the required channel capacity in GISC spectral camera with $30\%$ sampling rate in our experiment is $C_3\approx4.90\times10^5$. $C_3<C_2$ shows that GISC spectral camera has the higher information acquisition efficiency in a single exposure compared to the conventional camera.
With the development of optical imaging technology, many new imaging technologies (such as CT image\cite{hsieh2009computed}) are not based on the point-to-point imaging mode. However, because the correlation between pixels of image data doesn¡¯t be applied in the imaging reconstruction algorithm, the information acquisition efficiency of those new coding imaging technology also can¡¯t approaching the Shannon Limit determined by Information Theory for conventional optical imaging instruments. However, GISC imaging solution applies a spatial random phase modulation to satisfy the restricted isometry property (RIP)\cite{gong2012experimental} required by applying CS that makes the improvement of information acquisition efficiency of the imaging system possible. Comparing with CS imaging technology (such as Single-Pixel Imaging via Compressive Sampling\cite{duarte2008single}, coded aperture snapshot spectral imagers\cite{kittle2010multiframe}), which forces on the compressive sampling of electric signal after photoelectric conversion to improve the channel capacity utilization efficiency of the electric signal, GISC imaging solution improves the optical channel capacity utilization efficiency and achieves the compressive sampling of the image data during the imaging acquisition process, which opens the way of approaching the Shannon Limit determined by Information Theory in optical imaging instruments.
As a new optical imaging technology, GISC spectral camera provides a unique solution for the spectral imaging of dynamic processes. This GISC imaging solution may also be expanded to other multi-dimensional information (such as polarization information) acquisition\cite{morgan2003surface}, ultra-fast measurement\cite{gao2014single}, and super-resolution imaging\cite{wang2012quantum, donoho1992superresolution, candes2014towards}.

%\bibliography{bibfile3}



\end{document}